\documentclass{IEEEcsmag}

\usepackage[colorlinks,urlcolor=blue,linkcolor=blue,citecolor=blue]{hyperref}
\expandafter\def\expandafter\UrlBreaks\expandafter{\UrlBreaks\do\/\do\*\do\-\do\~\do\'\do\"\do\-}
\usepackage{upmath,color}
\usepackage{subfig}

\jvol{57}
\jnum{7}
\paper{8}
\jmonth{July}
\jname{Computer}
\jtitle{DDC: A Vision for a Disaggregated Datacenter}
\pubyear{2023}

\begin{document}

\sptitle{Article Type: Feature (Memory Disaggregation)}

\title{DDC: A Vision for a Disaggregated Datacenter}

\author{Mohammad Ewais}
\affil{University of Toronto, Toronto, ON, M5S 3G4, Canada}

\author{Paul Chow}
\affil{University of Toronto, Toronto, ON, M5S 3G4, Canada}

\markboth{FEATURE}{FEATURE}

\begin{abstract}\looseness-1Datacenters of today have maintained the same architecture for decades using the server as the primary building block. However, this traditional approach suffers from under-utilization of its resources, often caused by over-allocating these resources when deploying applications to accommodate worst-case scenarios. Specifically, servers can quickly drain their over-allocated memory resources while their CPUs are not fully utilized.

This problem gives rise to a different school of thought, where resources are disaggregated instead of tightly bound to servers. This can address the utilization problem by allowing each type of resource to be allocated, utilized and freed separately as required. New high performance communication protocols, like CXL, could pave the way for practical implementations of resource disaggregation. In this article, we argue it is time to reconsider the datacenter architecture as a whole. We present our vision for a disaggregated datacenter aided by well-established computer architecture design methodologies.
\end{abstract}

\maketitle

\chapteri{D}ifferent datacenter operators have reported resource under-utilization. For example, Amazon AWS reported very low CPU utilization ranging between 7\% to 17\%\cite{amazon_underutil}, while Google's datacenters reported a 28\% to 56\% CPU utilization\cite{google_underutil}, and Alibaba reported 20\% to 50\% CPU utilization for most of the time, coupled by 80\% to 100\%\cite{alibaba_underutil} memory utilization. 

These figures all point to a serious resource under-utilization issue in today's datacenters. This is mainly due to the way applications are deployed and the way they share a server's resources. To avoid swapping to disk, applications are allocated memory sufficient for their worst-case usage. As a result there can be an excess of CPU resources because not enough memory is available as evident in the aforementioned figures. This problem has an economic side effect, as it negatively affects both the purchase and operating costs of the datacenter. Specifically, a portion of the costs of these resources is going unused, and these resources actively drain power even while idle.

A natural solution for the under-utilization problem is to avoid the resource coupling altogether. If memory and CPU resources are allocated separately, putting aside the boundaries of a server, it becomes theoretically possible to allocate exactly the required amount of each without any utilization issues. Such resource disaggregation moves the datacenter away from a server-centric to a resource-centric model where each type of resource is logically-pooled together independent of other resources. Such resource disaggregation exists today in the context of storage, where multiple storage servers exist in a datacenter and are shared by many client nodes. The same idea has been explored in recent research for the context of memory, though these proposals have predominantly relied on intuition and ad hoc methods, making them challenging to compare. This article takes a top-down approach, leveraging analogies from traditional computer architecture. It systematically addresses all relevant aspects of memory disaggregation, presents a comprehensive model, and proposes a research platform that facilitates quantitative comparisons among different approaches.

\section{STATE OF THE ART}
\label{sec:art}

Resource disaggregation is a currently active topic with many contributions. A look at our recent survey of these works~\cite{survey} shows a wide spectrum of different proposals and implementations. These proposals mostly fall under two broad architecture categories, as shown in Figure~\ref{fig:art_state_sys}:
\begin{figure}[htp]
\centering

\subfloat[][Split]{%
  \includegraphics[clip,width=\columnwidth]{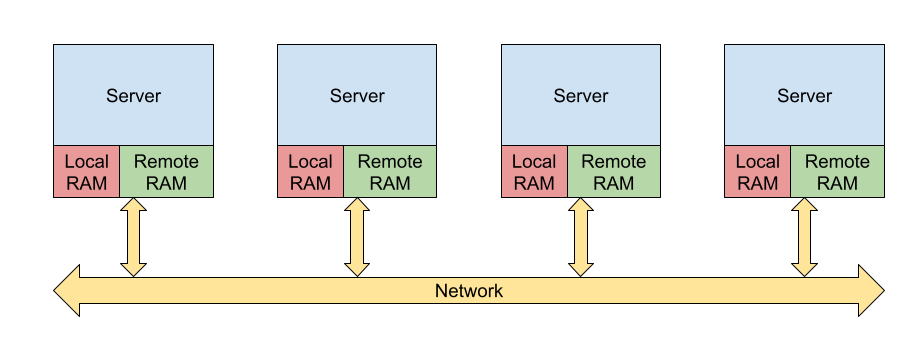}%
}

\subfloat[][Pool]{%
  \includegraphics[clip,width=\columnwidth]{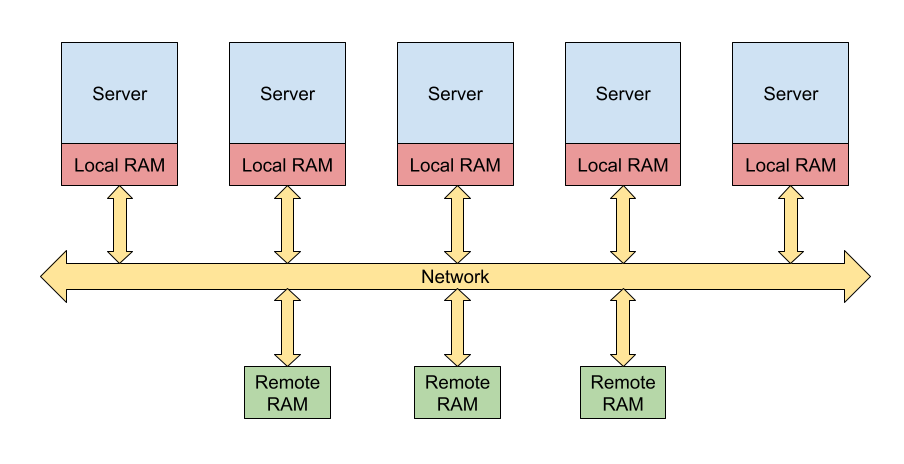}%
}

\caption{High-level system architectures for memory disaggregation}
\label{fig:art_state_sys}
\end{figure}
\begin{itemize}
    \item [{\ieeeguilsinglright}] \textbf{Split architecture:} Usually built using commodity servers. The server resources are split, where a partition of the resources is strictly used locally by the owning server, and the remainder of these resources are advertised for use by other nodes in the datacenter. Most of the implementations following this architecture are software, hypervisor, or OS based. 
    \item [{\ieeeguilsinglright}] \textbf{Pool architecture:} In this scheme, resources are more disaggregated by being pooled together. Typically, a cluster of this architecture would have dedicated memory nodes and compute nodes. The compute nodes still include some amount of memory that would act as a form of caching for the remote memory. Implementations of this architecture can be software, hypervisor, OS, or even hardware based.
\end{itemize}

\subsection{Limitations}
\label{sec:limits}
While almost all of the proposals studied by the survey follow one of the above mentioned general architectures, there are many specifics and details that vary by implementation. The survey also identified the following shortcomings of the state of the art:
\begin{itemize}
    \item [{\ieeeguilsinglright}] \textbf{Reasoning:} One important factor lacking in many of these studies is a reasoning as to the architectural choices made. Most architectural choices were based on intuition, personal opinion, or ease of implementation. This in turn has caused many of these studies to try and implement a fully functioning architecture in one step, instead of approaching this highly complex problem in a divide and conquer manner. Because of this, most studies only compare against a baseline of no disaggrgeation, but not against other disaggregation proposals.
    \item [{\ieeeguilsinglright}] \textbf{Ease of design and adoption:} A common trend in prior studies is the focus on ease of design at the expense of adoption. For example, many studies focus on software based implementations as they are the easiest to implement, but they force modifications or complete rewrites of existing applications. On the contrary, hardware based implementations can be the easiest to adopt (if they maintain backward compatibility) but are the hardest to design. We argue that this makes hardware a favorable choice since most datacenters are upgraded every three to five years~\cite{lifespan}, incurring that cost anyway.
    \item [{\ieeeguilsinglright}] \textbf{Latency:} One of the main constraints of a disaggregated system is latency, which comes from two main contributors. The first is software latency, which is incurred for using the various software network stacks, switching between different software layers (e.g., application to container to OS) with all the faulting and trapping required, etc. This component of latency can be mitigated by moving the implementation layer closer to hardware. The second major contributor to latency is the request fabric latency. This component of latency is inevitable, though it can be improved through the careful choice of network protocols, topology, and switching infrastructure. The emergence of a newer class of high throughput and low latency network protocols like CXL\cite{cxl} and RIFL\cite{rifl} promise major improvements to the latency of disaggregated systems.
    \item [{\ieeeguilsinglright}] \textbf{Data sharing:} Data sharing has largely been omitted by most recent studies, as supporting data sharing and coherence can significantly complicate any system design. In fact, as far as we know, the potential savings of allowing data sharing in disaggregated memory have not been studied or quantified. However, given that in a typical datacenter applications may be deployed in multiple replicated instances operating on the same datasets, allowing data sharing in disaggregated memory could prove useful.
\end{itemize}

\section{INSIGHT: REUSE THE WHEEL}              
In a disaggregated datacenter, there are five essential components. They are, compute units, memory, storage, interconnect, and a management layer to manage and control all these different components. Our intuition is then very simple. These are the same components that make up traditional computers, and while the scale is different, there is a clear analogy between a traditional computer and a disaggregated datacenter. 

\begin{figure}[t]
\centering
\includegraphics[width=\columnwidth]{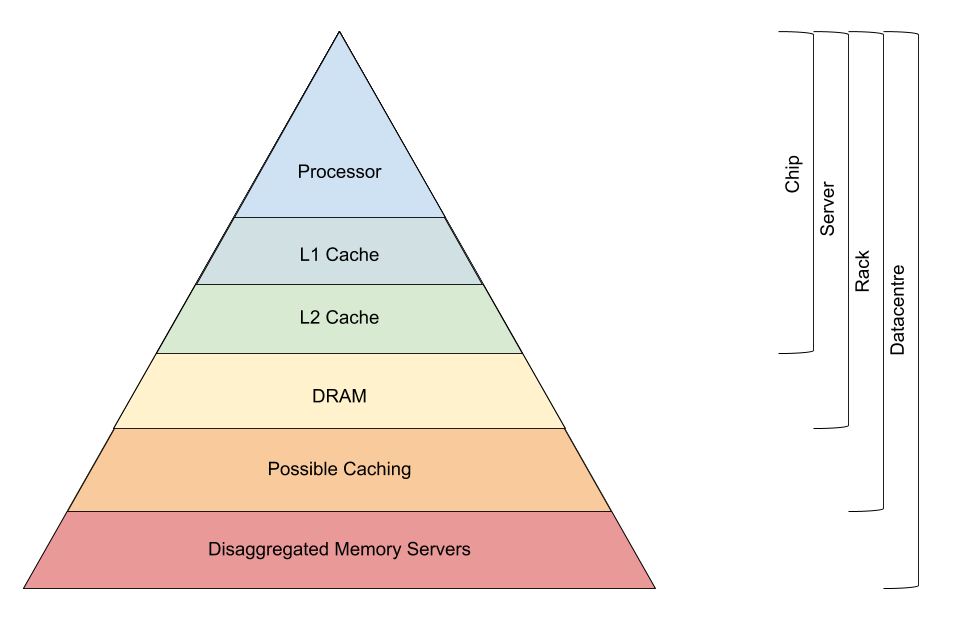}
\caption{Extended memory hierarchy in the datacenter}
\label{fig:mem_hier}
\end{figure}

The compute nodes in the datacenter can be thought of as the processors of the datacenter, while the remote memory may be analogous to the local DRAM available to any computer. The datacenter networks in this case serve as a processor interconnect (e.g., similar to on-chip buses). In this way of thinking, a disaggregated datacenter is similar to a traditional multi-core computer, and thus the same design philosophies of computers can also be utilized to build disaggregated datacenters. For example, introducing caching between the compute and the memory nodes could prove useful, as well as utilizing cache coherence protocols similar to the ones in microprocessors today, etc. In this case, the extra layers of memory disaggregation become a natural extension to the computer's memory hierarchy, as shown in Figure~\ref{fig:mem_hier}. 

For such an architecture to be of any practical use it must achieve the following requirements:
\begin{itemize}
    \item [{\ieeeguilsinglright}] \textbf{Ease of adoption:} For a fundamental change in datacenter architecture to be successfully adopted, it must achieve two things. First, datacenters run a wide array of different applications, from graph processing, databases, ML, to scientific computations. It is near impossible to reprogram all of these to utilize a new architecture. Instead, the architecture must provide backward compatibility. Second, the architecture must provide a straightforward programming model to ease the development of new applications. Based on our analogy above, the well-established thread-based shared memory model becomes the programming model of choice. Existing multi-threaded applications should, in theory, continue to function on such systems without modification, albeit on a much larger scale. 
    \item [{\ieeeguilsinglright}] \textbf{Performance:} Microprocessors today are designed to hide the latency of cache misses and long memory accesses. However, there is a limit to how much latency can be hidden before the the performance degrades. This is especially important in the case of disaggregated datacenters as the large scale puts extra pressure on the datacenter memory hierarchy and network interconnect to achieve low latencies and high bandwidths.
    \item [{\ieeeguilsinglright}] \textbf{Scalability:} Datacenters today are scalable, in the sense that more servers can be added or removed while the datacenter continues to function. A disaggregated datacenter must conform to the same requirements. However, in a disaggregated datacenter, there are two main types of scalability. The first is horizontal scalability that can be achieved by adding or removing compute components or memory components to/from the datacenter as it continues to function. The second type of scalability is vertical scalability, a capability non-existent in traditional datacenters. Vertical scalability is achieved by adding or removing layers of the memory hierarchy (e.g., extra levels of caching) to suit the datacenter needs, which can possibly be implemented dynamically through redefining the network interconnect.
    \item [{\ieeeguilsinglright}] \textbf{Fault tolerance:} Failures of nodes in traditional datacenters are a fairly common event, and must be handled properly. Furthermore, mass failures due to power outages or other reasons are also not uncommon. If fault tolerance is not supported or enabled, the results of such outages would be devastating. The same constraints would naturally apply to a disaggregated datacenter. It must be able to support fault tolerance actively, by introducing the capability of resuming applications on non-faulty hardware and within an acceptable delay, and passively, by allowing replication or data persistence to counter faults.
\end{itemize}

Based on these requirements and constraints, we believe that a hardware-based pool architecture should be the architecture of choice for a disaggregated datacenter. The performance requirements alone would probably disqualify non hardware-based implementations as they would suffer from higher latency. Also, a hardware-based implementation extending the memory hierarchy as discussed above would support a shared memory, multi-threading model and in turn support backward compatibility. Furthermore, from the general architectures discussed in the \nameref{sec:art} section only a pool architecture (as opposed to split) allows for better scalability both vertically and horizontally. It is also worth noting that we do not fully disqualify split architectures, as they can be thought of as special cases of pool architectures where the local part of memory is analogous to exclusive caches, and all the remote parts of memory simply form a NUMA. Thus, for the remainder of this article, and with this \textbf{hardware-based pool-like} architecture in mind, we approach its design from a computer architecture perspective and discuss the potential ways to design its components, and the design space parameters that researches need to look at.

\section{COMPUTE NODES}
In the context of disaggregated datacenters, compute nodes can simply be stripped down versions of today's servers. A compute node would only consist of its processor(s), local DDR memory, and a high speed network interconnect (e.g., a CXL interface) to extend the node with remote memory, but no further peripherals. The local DDR memory in this case does not function as a main memory, but rather as a big DRAM-based level of caching. Thus, all the fundamental changes to the known architecture of servers today will be around the DDR memory. For example, the memory controller will have to be modified to also function as a cache controller, handling replacements, evictions, etc. The memory/cache controller will also utilize the network interface to connect to the next level of caching, if any, or the remote memory.

Because the compute nodes are stripped down in our architecture, we expect them to be in the format of compact size blades, possibly allowing multiple of these to fit together in a backplane form. The idea of compute blades has been proposed before~\cite{dredbox} although utilized in a different architecture. This architecture of compute nodes need not be limited to CPUs, but may also be applicable for GPUs and other accelerators as well.

\section{CACHING}
\subsection{Caches}
In our envisioned architecture, there is at least one level of extra caching, which is the local DDRs of compute nodes. Whether extra levels of caches are needed depends on many factors, including the local DDR cache size and parameters, the interconnects and their topology, the scale of the disaggregated datacenter, etc. Designing these new levels of caching should, in principal, be no different from designing microprocessor caches, albeit on a much larger scale.

Outside the compute blades, we envision a level of caching at each granularity. For example, a backplane of compute blades could include one cache blade shared among the entire backplane. A second level of caching may exist in the rack, shared between a rack's compute backplanes, and so on. The placement of these caching levels is a function of the interconnects between these different granularities of compute resources, as well as the caches themselves, their sizes, bandwidth, etc. Regardless of the cache location or the granularity of compute resources it is attached to, all these caches would follow the same general high level architecture. The caches will consist of a DDR memory, along with a memory controller that has two dedicated interconnect interfaces going upward (towards compute blades) and downward (towards extra levels of caching or the memory nodes).

There are many design aspects of caches that are unknown such as, how to partition the memory itself to hold the necessary metadata and the directory, what cache line size is adequate, the replacement, insertion, and eviction policies of the cache, and the cache organization (e.g., associativity). We leave these to future work.

\subsection{Consistency and Coherence}
Since one of our goals is maintaining backward compatibility, the memory consistency model of our architecture will always be restricted to that used by the compute node processors. For example, x86 based systems should maintain a TSO consistency model, and similarly, a release consistency model should remain in effect for an ARMv8 based system, etc. Coherence, on the other hand may require some redesigning effort. To begin with, it is obvious that snooping based coherence is not suited for disaggregation, as it is known to not scale well. It also requires a lot of broadcasting which would be excessively strenuous on the interconnect, which is already a scarce resource. So, for our generalized architecture, directory based coherence becomes the approach of choice. The coherence protocol itself is left to be researched by future studies.

However, directory based coherence is not without its inefficiencies. Misses in a cache will result in a request to the parent directory, which in turn would send one or more requests to sharers to either invalidate or downgrade their data, followed by a response from one of these sharers to the original requester. This sequence of events means more pressure on the interconnect, as well as extra latency on every miss. To alleviate both of these drawbacks, prior studies~\cite{concordia,mind} have proposed to utilize the central locations of switches inside the interconnect, and used programmable switches to implement directories, essentially cutting these latencies in half. While programmable switches are limited in memory and thus can throttle or limit the size of the directory implemented, we believe an extension of such a methodology may be able to support larger solutions.

\section{MEMORY NODES}
The memory nodes in our envisioned architectures are dumb components, i.e., they have no active processing components. Instead, they would only consist of one or more DDR or HBM memory controllers, as well as one or more interconnect interfaces to the upper levels, and the necessary decoding and switching to map between the interconnect ports and the memory controllers. Similar ideas for memory blades have been proposed in \cite{blade,dredbox}. As an optimization, the memory nodes could be upgraded to support some compute units either inside or very close to the memory controllers to support a form of near memory computing. Fundamentally, memory nodes are no different from traditional main memories of servers today. The only main difference being that they sit behind a different class of interconnects.

\section{VIRTUAL MEMORY}
So far in our discussion we have omitted virtual memory. The natural inclination when extending the memory hierarchy with extra levels is to simply extend the virtual memory infrastructure (e.g., TLBs) with extra levels as well. However, this could prove inefficient. When a page fault happens, it first manifests as misses in each level of TLBs (typically, there is a TLB level for every caching level) all the way down until it reaches the memory blade that hosts the page table. Not only that, performing a page table walk (a search for an entry in the page table) at this stage would also incur the latency of accessing the remote memory as it will most likely not be cached. In short, without rethinking the design, even the infrequent event of page faults, or TLB misses can potentially take too long to be usable.

For example, a possible improvement would be to move page table walks from the processor side (where it is typically implemented) to memory nodes hosting the page table instead, where it can be trivially implemented. An opposite idea would be to host ``private'' page tables on the local DRAM of every compute node, which would cut down the trip needed on a page fault and make it comparable to that of traditional servers. Another approach would be to reconsider using virtual caches (caches that store data based on its virtual rather than physical address) along the hierarchy, though supporting coherence in virtual caches is harder, which makes them unfavorable beyond first level caches today. We leave the study of these and other potential solutions to future research.

\section{INTERCONNECTS AND NETWORKING}
Interconnects are arguably the most important component of a disaggregated datacenter. They not only affect the design choices of compute, caching, and memory blades, interconnects are also the deciding factor of the scale of disaggregation, i.e., the size of the disaggregated datacenter. In fact, without the advances made in networking today it would be impossible to achieve any usable form of disaggregation. 

Unfortunately, the requirements for networks, in terms of necessary bandwidth and latency, to support large scale disaggregation are unclear. Prior studies like \cite{network_support} and \cite{network_requirements} aimed at finding these requirements, but were very limited in terms of the underlying system used to collect such data. We believe it is time to revisit these requirements under more realistic conditions. Future research should investigate the network protocols, topologies, and switching infrastructure necessary to achieve disaggregation. This would determine the usability of emerging network protocols like CXL~\cite{cxl} and RIFL~\cite{rifl}, as well as outline the requirements of future research into networks, if needed.

\section{STORAGE}
In a traditional datacenter of today, the only disaggregated component is storage. This has been the case for many years now. Typically, each server would have some limited amount of storage, but there would also exist storage servers with lots of storage as well as duplication for fault tolerance, which would be shared by the entire datacenter. This already integrates well with our philosophy, adding an extra layer of the memory hierarchy beyond the remote memory.




However, disaggregation presents a unique opportunity to consider a different storage architecture. Namely, it may be beneficial to consider a unified storage and memory hierarchy, or in other words, a unified volatile and persistent memory space. While some prior studies~\cite{hotpot} have proposed using non-volatile memories as remote memories, we believe adapting the memory blades discussed previously to also work with storage would suit our proposed architecture better. Such unified memory-storage hierarchies can avoid the unnecessary moving of pages between storage and memory. Instead, pages that are more frequently used can reside in the faster memory, while those less frequently accessed may reside in storage or non volatile memories instead.

However, a problem still remains, which is how to address this unified memory from a user perspective. Memories are byte addressable, while storage typically requires a method of naming data (e.g., directories and files). In a unified hierarchy it is not immediately clear what method should be used to not break backward compatibility. We leave the exploration and decision of such aspects to future work.

\section{DESIGNING AND PROTOTYPING}
Computer architecture and memory hierarchy design are often aided by the use of simulators, which allow for the manipulation and exploration of different design parameters as well as testing new/different architectures or organizations. Since the components we have described so far are very similar to their counterparts of traditional computers, we suggest the use of traditional computer architecture simulators for the exploration and design of memory disaggregation systems as well, though these simulators would require modifications/expansions to handle such large scales. The use of such simulators would also provide a standard methodology for comparing and evaluating different designs, a trait fundamentally needed in the field of memory disaggregation.

Prototyping such designs would also be possible through the use of commercially available FPGA boards, many of which include two 100Gbps interfaces as well as DDR4/HBM and NVMe that would make them suitable for prototyping cache, memory, and storage nodes. Variants of such boards that include MPSoC FPGAs (FPGAs tightly coupled with CPU cores) would be suitable for prototyping compute nodes as well. In fact, it may even be possible to use FPGAs for final implementations, not only prototyping, which would open the door to utilize their field programmability in interesting ways, as we describe later.

\section{MANAGEMENT}
\begin{figure*}[htp]
\centering
\includegraphics[width=\textwidth]{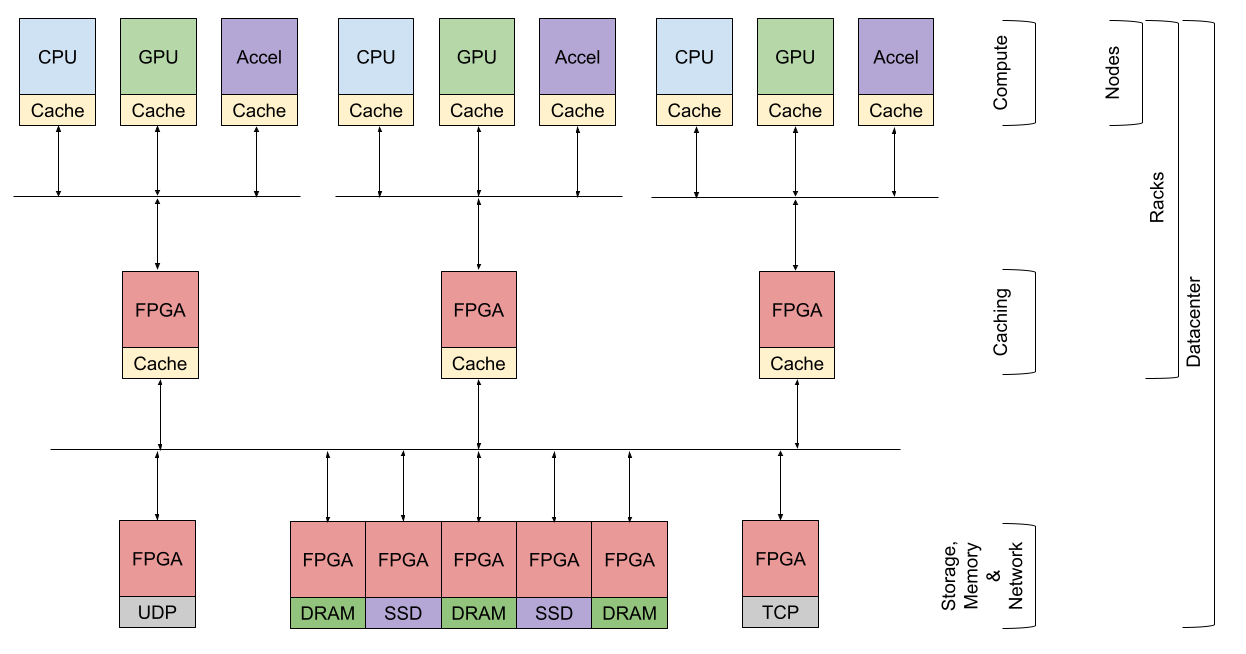}
\caption{An example of our proposed architecture. The upper layer represents heterogeneous compute nodes, with their local DRAMs functioning as caches. The middle layer represents one level of FPGA based caching. The bottom layer represents a hybrid remote memory/storage as well as NICs interfacing with the outside world. The interconnect is simplified and does not show the topology nor the switching infrastructure also serving as directories.}
\label{fig:example_arch}
\end{figure*}
Figure~\ref{fig:example_arch} shows an example of our proposed architecture. The architecture includes different types of compute nodes with their local DRAM caches. The following layer is one or more levels of cache blades that could exist on different levels of grouping (e.g., a rack). The last part of our architecture is a mix of storage and memory blades, as well as network interface cards to allow communicating with the outside world.

With our established architecture, one last crucial component of its design remains. Namely, a management layer. Because our system closely resembles that of a traditional computer, using an unmodified operating system would be theoretically possible, although it would be far from optimal in managing these resources. We believe some modifications to a traditional OS to be necessary, and some others may be useful as an improvement:
\subsection{Resource Monitoring}
In a traditional computer, the operating system builds a list of the connected devices and the memory map once at boot time. Since there is no risk of partial system failure, there is also no need for continuous monitoring of the resources or updating the list of memory/resources. In later years, OSes started supporting a limited form of dynamic resource addition/removal to serve virtualization, such as the Linux HotPlug technology\cite{hotplug}.

However, in a datacenter, an extension of this caliber is not enough. Rather, it must be in the nature of the operating system itself to expect a dynamically changing system, in terms of memory as well as compute resources. This is necessary for many reasons. First, a datacenter operator must be able to scale up or down their use of the datacenter as needed. Some resources may be powered down or put to sleep when unneeded. This is crucial as power consumption is one of the important factors of operating datacenters. Second, datacenters often suffer from device failure, which must be mitigated with minimal delay to not affect their tasks. Such failures could occur at any type of resources, and can be alleviated in a variety of ways the discussion of which is left to later studies. However, a datacenter operating system must be ready and able to handle such issues shall they arise.

To address these issues, we adopt a similar design to LegoOS~\cite{lego_os}. In short, some compute nodes will function as hosts for the OS itself, which must also be replicated to avoid a single point of failure. Every type of resource will run a monitor (either in software or in hardware) that will periodically send its status and usage statistics to the OS nodes. The OS nodes are then responsible for servicing all requests for application deployment, memory allocations, as well as recovering from hardware faults, taking into account the periodic status collected. To unburden the OS from doing these tasks on a low level scale, which may require extra unnecessary communication between the OS and the target nodes, the OS should only be responsible for such deployments and allocations on a macro scale. For example, the OS may decide that an application can be deployed on two neighboring compute blades, and will have its memory allocated in a certain memory blade. But it does not need to handle the specific address in the memory blade where memory will be allocated, or the specific thread affinity in the compute blades. Such micro-level details can be left to the monitors running on each of these blades. This makes the OS handling of resource allocation somewhat hierarchical.

After changing or applying these modifications to the OS Kernel itself, the remainder of the OS can continue working unmodified, simply viewing the datacenter-scale computer as a huge traditional computer. Its existing algorithms for scheduling, allocation, etc., may be used as is. Whether these algorithms are suitable for use at such large scale or may require modifications to perform better is left to future work.

\subsection{Software Defined Architecture}
FPGAs of today are equipped with 100Gbps interfaces, and will soon be enabled with 400Gbps. Together with their parallelism, they are not only suitable for prototyping, but potentially full implementations of our components. This presents us with an unprecedented opportunity for improvement, which is programmability.

Simply put, if caches are built using FPGA boards, then they can have their traits and features modified at run time. For example, we may decide that an existing replacement policy is not performing well with a certain type of application, and can instead be upgraded to another replacement policy that would work better. More fundamental changes can occur, such as changing the cache organization, associativity, etc. Such features can only be enabled by having FPGA-based implementations. In the case of caches (but not necessarily other nodes), this may require exposing the caches to the OS and sending statistics to it.

The same concept can be applied on a macro level, where the interconnects of the datacenter are basically networks. If the switching devices for these networks are FPGA based, or at the very least are programmable switches, then we can also dynamically modify the architecture itself at run time. Entire layers of caching may be exposed or hidden, FPGA blades can completely have their functions modified, e.g., from cache to memory node, or vice versa, etc. The only unmodifiable components in this case are the compute blades, apart from which we can have a software-defined memory hierarchy.

\section{CONCLUSION}
The article presents a top-down approach to designing a disaggregated memory datacenter in a way that is easy to adopt, maintains backward compatibility, and scalability by looking at the datacenter itself as a single large scale computer. We discuss the possible designs for the components of such a computer as well as how to manage it, and present a research methodology that facilitates quantitative analysis and comparison. We hope this proves insightful for future research into the area.\vspace*{-8pt}

\def\refname{REFERENCES}

\vspace*{-8pt}

\begin{IEEEbiography}{Mohammad Ewais}{\,} received his master's degree in computer engineering from the University of British Columbia. He is currently pursuing his PhD in computer engineering at the University of Toronto. His main research interests include computer architecture, datacenter architecture, reconfigurable computing and FPGAs. Email: \url{mewais@ece.utoronto.ca}.\vadjust{\vfill\pagebreak}
\end{IEEEbiography}

\begin{IEEEbiography}{Paul Chow}{\,} is a professor in the Electrical and Computer Engineering department at the University of Toronto. His main research focuses on reconfigurable computing with an emphasis on programming models, middleware to support programming and portability, and scaling to large-scale, distributed FPGA deployments. Email: \url{pc@eecg.toronto.edu}.\vspace*{8pt}
\end{IEEEbiography}

\end{document}